\documentclass[useAMS,usenatbib]{mn2e}
\usepackage{graphicx}

\title[]{On the possible mechanism to form the radio emission
spectrum of the Crab pulsar}

\author[Machabeli G. and Chkheidze N.]{Machabeli G.\thanks{E-mail:
g.machabeli@iliauni.edu.ge} Chkheidze N. \\
Centre for Theoretical Astrophysics, ITP, Ilia State University, 0162-Tbilisi, Georgia\\
}
\begin{document}

\date{Accepted 2014 March 07}

\pagerange{\pageref{firstpage}--\pageref{lastpage}} \pubyear{2009}

\maketitle

\label{firstpage}

\begin{abstract}
In the present paper a self-consistent theory, explaining shape of
the observed phase-averaged radio spectrum in the frequency range
from 100MHz to 10GHz is presented. The radio waves are assumed to be
generated near the light cylinder through the cyclotron resonance.
The cyclotron instability provides excitement of the
electron-positron plasma eigen-waves, which come in radio domain
when the resonant particles are the most energetic primary beam
electrons. It is widely accepted that the distribution function of
relativistic particles is one-dimensional at the pulsar surface. The
generated waves react back on the resonant particles causing their
diffusion in the perpendicular direction to the magnetic field and
violating the one-dimensionality, which switches on the synchrotron
radiation process. The synchrotron emission of the beam electrons
provides generation of high-energy $\gamma$-rays simultaneously with
the radio emission, that explains the observed pulse
phase-coincidence in these energy domains. The theory provides a
power-law radio spectrum with the spectral index equal to $3.7$,
coming in agreement with the observations.
\end{abstract}


\section{Introduction}

The only energy source providing the observed emission of Crab
nebula and pulsar, is the pulsar rotation slowdown energy. Thus, one
of the main problems in pulsar emission theory is to find the
transformation mechanism of the rotational energy into the observed
radiation. Due to works \citet{gold,stur} and etc. a well-defined
scheme of serial processes has developed, which provides the
solution of the mentioned problem. In particular, we believe that at
the polar cap due to rotation of the magnetized neutron star the
electric field is generated, which extracts the primary beam of
electrons from the pulsar surface and accelerates them. Moving along
the weakly curved magnetic field lines the beam electrons start to
generate $\gamma$-quanta. As long as the energy of gamma-quanta in
the laboratory frame satisfies the condition
$\varepsilon_{\gamma}\sin\alpha>2m c^{2}$ (here $\alpha$ is the
angle between the magnetic field and the direction of motion of the
$\gamma$-quanta) in the strong pulsar magnetic field develop the
quantum effects of cascade creation of electron-positron
($e^{-}e^{+}$) pairs \citep{ru1,mich}. The size of the region near
the star surface where the longitudinal component of the electric
field is nonzero (gap region) does not exceed few meters. After
leaving the gap region charged particles move along the magnetic
field lines, as they loose their perpendicular momenta in a very
short time ($\leq10^{-20}$s) due to the synchrotron radiation
processes. Thereby, the magnetospheric $e^{-}e^{+}$ plasma is
formed, with an anisotropic one-dimensional distribution function
(see Fig.1 from \citet{aro}), in which generate the electro-magnetic
waves leaving the pulsar magnetosphere and reaching the observer as
the pulsar emission.

The Crab pulsar differs from the most of radio pulsars, as emits in
a broad range, from radio up to very high energy gamma-rays. The
important observational feature of the Crab pulsar emission is, that
its multiwavelength radiation pulses are coincident in phase
\citep{manch,veritas}. This feature should indicate that the
emission from different energy bands is generated simultaneously at
the same location in the pulsar magnetosphere. In our previous works
\citep{ch10,chcrab} was considered the simultaneous generation of
the high-energy $\gamma$-rays from $10$MeV to $400$GeV, and the
radio waves at the light cylinder length-scales in the magnetosphere
of the Crab pulsar. The reason for the wave generation in the outer
parts of the pulsar magnetosphere is the one-dimensionality and
anisotropy of the particles' distribution, which leads to
development of plasma instabilities. In particular, in the Crab
pulsar's magnetosphere near the light cylinder the conditions for
the cyclotron resonance are satisfied, which provides generation of
the low frequency cyclotron waves \citep{kmm}. The resonant
particles diffuse in both directions along and across the magnetic
field during the quasi-linear relaxation stage of the instability
and acquire the perpendicular momenta. Consequently, the particles
start to radiate through the synchrotron mechanism. We assumed that
the resonant particles are the most energetic primary beam electrons
with the Lorentz factors $\gamma\sim10^{7}-10^{9}$, which provides
the synchrotron radiation of energetic $\gamma$-rays up to $400$GeV.
Such high energies by beam electrons are reached during their
re-acceleration in the light cylinder region due to the Landau
damping of electrostatic waves (see \citet{maho,chcrab} for
details). At the same time such particles can provide generation of
the cyclotron modes in the radio domain. In previous work
\citep{chcrab} we only considered the theoretical explanation of the
measured $\gamma$-ray spectrum of the Crab pulsar and explained
coincidence of $\gamma$-ray and radio signals.

Consequently, in the present paper we consider generation of radio
waves via the cyclotron resonance in more details (Sect. 2),
calculate the theoretical radio-spectrum (Sect. 3) and make
conclusions (Sect. 4).

\section{Generation of radio emission} \label{sec:consid}

Differently from gamma-ray emission, which does not 'feel' the
medium and emerges freely from the region of its generation, for the
low frequency radiation the condition is satisfied $\lambda\gg
n^{-1/3}$ in the pulsar magnetosphere (here $n$ is the density of
plasma particles, and $\lambda$ is the wave-length of low frequency
waves). In this case one should take into account the effects of
wave interference, which means applying the plasma physics methods.
Such problem was considered for the cold plasma in hydrodynamical
approach in works \citep{bu, harde}. And for the plasma with an
arbitrary temperature this problem was solved in kinetic approach,
in the following works \citet{volo,arons,phd}.

It is usually suggested that the pulsar magnetosphere is filled by
relativistic  $e^{-}e^{+}$ plasma. In strongly magnetized cold
$e^{-}e^{+}$ plasma can be excited only waves propagating almost
along the magnetic field lines \citep{kazo}. As we have already
mentioned during the pair creation near the star surface, the
perpendicular momenta of charged particles are instantly lost
through the synchrotron radiation, making the particle distribution
one-dimensional. The plasma with such distribution is unstable and
excites plasma eigen-waves (the so called $t$-waves) at a certain
resonance condition \citep{lolo}. The $t$-wave is a purely
transverse wave with an electric field perpendicular to
$(\mathbf{k},\mathbf{B})$ plane, where $\mathbf{k}$ is the wave
vector and $\mathbf{B}$ is the vector of the magnetic field. The
dispersion relation of the $t$-mode is written as
\begin{eqnarray}\label{}
    \omega^{t}=kc(1-\delta),
\end{eqnarray}
where $\delta=\omega_{p}^{2}/(4\omega_{B}^{2}\gamma_{p}^{3})$,
$\omega_{B}=eB/mc$ is the cyclotron frequency and the square of the
plasma frequency $\omega_{p}^{2}=4\pi n_{p}e^{2}/m$.

The resonance condition of the cyclotron instability which can
provide generation of $t$-waves can be easily obtained, if
considering the difference between the quantum synchrotron levels
$l$ and $f$ \citep{jele,ginz}
\begin{equation}\label{}
    E_{l}-E_{f}=\left(m^{2}c^{4}+p_{l}^{2}c^{2}\right)^{1/2}-\left(m^{2}c^{4}+p_{f}^{2}c^{2}\right)^{1/2}=\hbar\omega_{lf},
\end{equation}
where $p^{2}=p_{\parallel}^{2}+p_{\perp}^{2}$. Taking into account
that $m^{2}c^{2}/p^{2}\approx1/\gamma^{2}\ll1$ and
$p_{\perp}^{2}/p_{\parallel}^{2}\ll1$ are the small parameters, one
can expand Eq. (2) in series and obtain
\begin{equation}\label{}
    \left(p_{\parallel_{l}}-p_{\parallel_{f}}\right)\upsilon_{\parallel}+\frac{1}{2m
    \gamma}\left(p_{\perp_{l}}^{2}-p_{\perp_{f}}^{2}\right)\approx\hbar\omega_{lf}.
\end{equation}
Assuming, that $p_{\parallel_{l}}-p_{\parallel_{f}}=\hbar
k_{\parallel}$ and
$\left(p_{\perp_{l}}^{2}-p_{\perp_{f}}^{2}\right)/(2m)=\hbar
s\omega_{B}$, where $s\equiv l-f$ denotes the difference between the
synchrotron landau levels $l$ and $f$, we will get
\begin{equation}\label{}
    \omega_{r}-k_{\parallel}\upsilon_{\parallel}-s\frac{\omega_{B}}{\gamma}=0.
\end{equation}
It is clear that, when $s=-1$ (the anomalous Doppler effect) the
wave energy grows at the expense of longitudinal energy
$\omega_{r}k_{\parallel}\upsilon_{\parallel}<0$, i.e.
$\upsilon_{\parallel}>\upsilon_{ph}$ (here $\upsilon_{ph}=\omega/k$
is the phase velocity of excited waves) and waves gain particles
energy through the interaction process. The waves, as well as the
particles leave the pulsar magnetosphere in time of the order of
$\Omega^{-1}\approx10^{-2}$s (here $\Omega=2\pi/P$, and $P$ is the
pulsar spin period). This does not prevent the instability process,
as the resonance area is always filled with newly arrived particles
which take the place of particles that leave this region. The
frequency of excited $t$-waves can be estimated from Eq. (4) as
follows \citep{chcrab}
\begin{equation}\label{}
    \omega_{r}\approx\frac{\omega_{B}}{\delta\gamma}.
\end{equation}
As we see the frequency of generated $t$-waves depends on the
Lorentz-factors of the resonant particles, the higher is the energy
of the particles the lower is the frequency of the excited
eigen-waves. For the most energetic particles in the pulsar
magnetosphere, the primary beam electrons (which Lorentz factors
vary from $10^{7}$ to $10^{9}$, for PSR B0531+21, as assumed in
\citet{chcrab}) the frequency of the transverse waves come in the
radio domain. In particular, it varies in the frequency range from
$100$MHz to $10$GHz.

Under influence of the electric field of the generated radio waves
the resonant particles diffuse along and across the magnetic field
lines \citep{kazo}. This causes the particles' redistribution and
the distribution function acquires the perpendicular component by
impulses. The growth of the perpendicular impulses continue till the
balance is achieved between the processes of pumping the
longitudinal energy of particles into their transversal motion and
the energy losses caused due to their synchrotron radiation. This
process can be described by the following equation in kinetic
approach (see Eq. (11) in \citet{chcrab})
\begin{equation}\label{}
    \frac{\partial \textit{f
}^{0}}{\partial
p_{\perp}}=\frac{F_{\perp}}{D_{\perp,\perp}}\textit{f }^{0}.
\end{equation}
where $D_{\perp,\perp}=e^{2}/(8c)\delta|E_{k}|^{2}_{k=k_{res}} $ is
the transversal diffusion coefficient, $|E_{k}|^{2}$ is the density
of electric energy in the excited waves and $F_{\perp}$ is the
transversal component of the synchrotron radiation reaction force,
which has the form
\begin{equation}\label{}
    F_{\perp}=-\alpha_{s}\frac{p_{\perp}}{p_{\parallel}}\left(1+\frac{p_{\perp}^{2}}{m^{2}c^{2}}\right),
    \qquad\alpha_{s}=\frac{2e^{2}\omega_{B}^{2}}{3c^{2}}.
\end{equation}
From Eq. (6) one can easily find
\begin{equation}\label{9}
    \textit{f}(p_{\perp})\propto
    e^{-\left(\frac{p_{\perp}}{p_{\perp_{0}}}\right)^{4}},\qquad p_{\perp_{0}}=\left(\frac{4\gamma_{r}m^{3}c^{3}}{\alpha_{s}}D_{\perp,\perp}
    \right)^{1/4}.
\end{equation}
During the quasi-linear stage of the resonance the excited waves
cause particle diffusion as along also across the magnetic field
lines. The transversal diffusion leads to appearance of the
perpendicular impulses and the one-dimensional distribution function
acquires the transversal component. The particles emit in the
synchrotron regime when having the perpendicular component of the
impulse. The characteristic frequency of the synchrotron radiation
of a single electron is
$\epsilon_{syn}\approx5\cdot10^{-18}B\psi\gamma^{2}$GeV
\citep{ginz}. The energy of the synchrotron photon reaches
$\epsilon_{syn}\approx400$GeV, if the Lorentz factor of the emitting
particles is of the order of $10^{9}$. The particles' acceleration
mechanism up to Lorentz factors of the order of $10^{9}$ is
considered in previous work \citet{chcrab}.

The synchrotron radiation reaction force plays the main role in the
process of wave generation and redistribution of the resonant
particles. In particular, the transversal component of the
synchrotron emission force $F_{\perp}$ confines the growth of the
pitch angles and the longitudinal component
$F_{\parallel}=-\alpha_{s}\gamma_{r}^{2}\psi^{2}$ redistributes the
emitting particles by their parallel momenta (this force appears to
be stronger than the diffusion forces). The longitudinal
distribution function of the resonant particles after the
quasi-linear relaxation takes the following form (see Eq. (27) in
\citet{chcrab})
\begin{equation}\label{}
    \textit{f}_{\parallel}\propto\frac{1}{p_{\parallel}^{1/2}|E_{k}|}.
\end{equation}
The spectral density of the excited waves $|E_{k}|^{2}$ depends on
the initial distribution function of the emitting particles (for the
initial moment is chosen the moment when the wave excitation
starts). At the initial moment the distribution of the beam
particles can be presented as \citep{chcrab}
\begin{eqnarray}\label{}
    \textit{f}_{\parallel_{0}}\propto\left\{%
\begin{array}{c}
  p_{\parallel}^{-n}, \;\; p_{\parallel _{min}}\leq p_{\parallel}\leq p_{\parallel _{s}}\\
  p_{\parallel}^{-m}, \;\; p_{\parallel _{s}}\leq p_{\parallel}\leq p_{\parallel _{max}}   \\
\end{array}%
\right\}.
\end{eqnarray}
Here $p_{\parallel _{min}}=mc^{2}\gamma_{b_{min}}$, $p_{\parallel
_{s}}=mc^{2}\gamma_{b_{s}}$, $ p_{\parallel
_{max}}=mc^{2}\gamma_{b_{max}}$ and it is assumed that
$\gamma_{b_{min}}\simeq6\cdot10^{6}$,
$\gamma_{b_{s}}\simeq4\cdot10^{7}$ and
$\gamma_{b_{max}}\simeq10^{9}$. One can easily represent the
function (10) by the following expression
\begin{eqnarray}\label{}
    \textit{f}_{\parallel_{0}}\propto
    p_{\parallel}^{-n}+\beta p_{\parallel}^{-m},
\end{eqnarray}
where $\beta\simeq1.3\cdot10^{-10}$ defines the break point of the
initial distribution between two power-laws. The power-law with the
index $m$ corresponds to re-accelerated beam particles near the
light cylinder via the Landau damping (see \citet{chcrab} for
details). Referring to the Eq. (36) from \citet{chcrab}
\begin{equation}\label{}
   a\frac{\partial}{\partial
    p_{\parallel}}\left(\frac{|E_{k}|}{p_{\parallel}^{1/2}}\right)+\textit{f}_{\parallel_{0}}=0,
\end{equation}
one can find the energy density of the cyclotron waves using the Eq.
(11) for $\textit{f}_{\parallel_{0}}$, which writes as
\begin{eqnarray}\label{}
    |E_{k}|\propto p_{\parallel}^{3/2-n}+\beta\frac{n-1}{m-1} p_{\parallel}^{3/2-m}.
\end{eqnarray}
Thus, for the longitudinal distribution function of the emitting
particles after the stationary state is reached from Eq. (9) we will
get
\begin{eqnarray}\label{}
    \textit{f}_{\parallel}\propto \left(p_{\parallel}^{2-n}+\beta\frac{n-1}{m-1} p_{\parallel}^{2-m}.\right)^{-1}.
\end{eqnarray}
The redistribution of the resonant particles via the synchrotron
radiation reaction force influences the process of excitement of the
$t$-waves. As the waves are generated on account of the longitudinal
energy of particles with the distribution function given by Eq.
(14).

\section{The spectrum of the radio emission}

For easy understanding of the formation of spectrum of excited
$t$-waves we use the quantum mechanical approach. Let us consider
two energy levels $l$ and $f$ ($l>f$). The number of particles at
$l$ level is $N_{l}$ and $N_{f}$ is the number of particles at the
energy level $f$. Through the decay of particles from a higher
energy level ($l$) to a lower one ($f$) the radio waves are
generated. The process can be described by the coefficient
$A^{l}_{f}$, which gives the probability per unit time that the
particle in state $l$ will decay to state $f$. The quantity
$A^{l}_{f} N_{l}$ in this case shows the mean value of transitions
from state $l$ to state $f$ per unit time. At the same time
particles fill the energy level $l$, as they interact with the
electromagnetic field of the waves with the frequency of the
transition between energy levels $l$ and $f$. It should be mentioned
that particles which provide the generation of the observed radio
emission also take place in the synchrotron emission process, as
through the cyclotron resonance they acquire pitch angles. The
synchrotron mechanism generates high-frequency radiation that freely
leaves the magnetosphere. The only influence of this process on the
excitement of radio waves is shown via the synchrotron radiation
reaction force that provides redistribution of the resonant
particles (see Eq. (14)), which interact with electromagnetic field
of radio waves and fill the energy level $l$. The mean value of such
transitions can be given by the following quantity $B^{f}_{l}
N_{f}u(\nu)$, where $u(\nu)$ is the spectral energy density of the
radiation field at the frequency of transition. The coefficient
$B^{f}_{l}$ is the probability per unit time per unit spectral
energy density of the radiation field that particle at state $f$
will jump to state $l$. The difference between the number of
particles of populated energy levels is constantly maintained, due
to arrival of the new particles in place of ones leaving the pulsar
magnetosphere. Consequently, the energy loss should not be taken
into account. Now if we equal the numbers of transitions between
energy levels $l$ and $f$ that define the radio domain by writing
\begin{equation}\label{}
    A^{l}_{f} N_{l}=B^{f}_{l} N_{f}u(\nu),
\end{equation}
we will obtain the expression for the spectral energy density of the
radio emission, which writes as
\begin{equation}\label{}
      u(\nu)=\frac{A^{l}_{f}}{B^{f}_{l} }\frac{N_{l}}{N_{f}}.
\end{equation}
Here we have not taken into account the stimulated transition of
particles from level $l$ to $f$, as it does not play a role in
radio-wave generation. Such transitions correspond to generation of
$\gamma$-rays through the synchrotron radiation. The ratio of
probabilities of photon emission $A^{l}_{f}$  and absorption
$B^{f}_{l} $ for any pair of levels $l$ and $f$ can be described by
$|E_{k}|^{2}$, which can be easily found from Eq. (13). And the
relation between the transition numbers $N_{l}/N_{f}$ is given by
Eq. (14). If we use the following expression $p_{\parallel}=\hbar k$
in place of Eq. (16) we can write
\begin{equation}\label{}
    u(k)\propto\frac{\left(k^{3/2-n}+\beta(n-1)/(m-1)k^{3/2-m}\right)^{2}}{k^{2-n}+\beta(n-1)/(m-1)k^{2-m}}.
\end{equation}
The shape of the radio spectrum can be defined after we give values
for $n$ and $m$. In previous work \citep{chcrab}, we showed that for
explaining the observed $\gamma$-ray spectrum of the Crab pulsar up
to $400$GeV one should assume that $n=6$ and $m=4.7$. In the
frequency range from $100$MHz to $10$GHz, where the Crab pulsar's
radio emission is detected the model spectrum defined by Eq. (16)
behaves like $u(\nu)\propto \nu^{-3.7}$ for the above given values
of $n$ and $m$. This value is close to average value of the measured
radio spectral index of the Crab pulsar in the given frequency
range, which we define as $(\alpha_{MP}+\alpha_{IP})/2\approx-3.6$,
where the values of spectral indices for main pulse and interpulse
are $\alpha_{MP}=-3.0$ and $\alpha_{IP}=-4.1$ (see Fig. 6 from
\citet{moffet}).

\section{Conclusions}

The emission generation mechanism in the outer parts of pulsar
magnetosphere suggested in the present work provides explanation of
the interesting observational feature of the Crab pulsar, the
pulse-phase alignment of $\gamma$-ray signals in the energy domain
$(0.01-400)$GeV with the radio signals. This behaviour is provided
due to simultaneous generation of emission in the high and the low
frequency ranges at the same location in the pulsar magnetosphere.
To assure the validity of the proposed emission model, one should
also explain the measured radio and $\gamma$-ray spectra. In work
\citet{chcrab}, we calculated the theoretical spectrum of the Crab
pulsar in the $10$MeV to $400$GeV energy domain, which was presented
by the following function
$F(\nu)\propto\left[\left(\nu/\nu_{0}\right)^{-(n-2)/(n-4)}+\left(\nu/\nu_{0}\right)^{-(m-2)/(m-4)}\right]^{-1}$,
where $\nu_{0}\approx9.7\cdot10^{23}$Hz (this corresponds to photon
energy $\epsilon_{0}\approx4$GeV and shows the break point on the
spectrum). To match the theoretical spectrum with the measured one
best described by a broken power-law of the form
$\left[\left(\nu/\nu_{0}\right)^{-1.96}+\left(\nu/\nu_{0}\right)^{-3.52}\right]^{-1}$
\citep{veritas, alek2}, we assumed that $n=6$ and $m=4.7$. In
present work we obtained the theoretical radio spectrum of the
following form (see Eq. (17))
\begin{eqnarray}\label{}
    F(\nu)\propto\frac{\left(\left(2\pi/\lambda\right)^{(m-n)}\nu^{3/2-n}+\beta(n-1)/(m-1)\nu^{3/2-m}\right)^{2}}{\left(2\pi/\lambda\right)^{(m-n)}\nu^{2-n}+\beta(n-1)/(m-1)\nu^{2-m}}.\nonumber
\end{eqnarray}
After using the values for $n$ and $m$ defined from the high energy
spectrum, we found that in the radio domain, from $100$MHz to
$10$GHz, the spectral function shows the simple power-law behaviour
$F(\nu)\propto\nu^{-3.7}$.

The theoretically inferred value of the spectral index is close to
the index of the radio emission phase-averaged spectrum, which
equals 3.6. It should be mentioned that the radio spectral indices
for the main and interpulse differ. Particularly, index of the main
pulse equals to 3.0, when the interpulse reveals the steeper radio
spectrum with the index equal to 4.1 \citep{moffet}. This
observational fact might be caused by different reasons. One of the
possible explanations could be the difference in distribution
functions of particles responsible for the radio emission
generation. This is very likely if the Crab pulsar is an orthogonal
rotator, then the main and interpulse should be emitted from
different poles of the star. Generation of particle beam with
exactly the same distribution function in this case would be
practically impossible.

It would be interesting to estimate the power of the radio emission.
For this one can write the condition for conservation of quanta
number, which has the following form
\begin{equation}\label{}
        \frac{|E_{k}|^{2}}{\omega_{k}}=const.
\end{equation}
Consequently, for the relation of gamma-ray and radio emission wave
energy densities one can write
\begin{equation}\label{}
        \frac{|E_{k}|^{2}_{\gamma}}{|E_{k}|^{2}_{r}}=\frac{\omega_{\gamma}}{\omega_{r}}.
\end{equation}
After substituting the values for radio and gamma frequencies from
Eq. (5) and Eq. (17) from \citep{chcrab}, one finds that
$\omega_{\gamma}/\omega_{r}\simeq\left(10^{6}-10^{8}\right)$. Taking
into account that the measured power of high energy emission of the
Crab pulsar is of the order of $10^{36}$erg/s, one can estimate the
approximate power for the radio emission which from Eq.(19) follows
to be of the order of $10^{28}-10^{30}$ coming in a good agreement
with the measurements.

\section*{Acknowledgments}

The research of the authors was supported by the Shota Rustaveli
National Science Foundation grant (N31/49).


\label{lastpage}

\end{document}